\documentclass{SciPost}

\hypersetup{
    colorlinks,
    linkcolor={red!50!black},
    citecolor={blue!50!black},
    urlcolor={blue!80!black}
}

\usepackage[bitstream-charter]{mathdesign}
\usepackage{bm}
\DeclareSymbolFont{usualmathcal}{OMS}{cmsy}{m}{n}
\DeclareSymbolFontAlphabet{\mathcal}{usualmathcal}

\fancypagestyle{SPstyle}{
\fancyhf{}
\lhead{\colorbox{scipostblue}{\bf \color{white} ~SciPost Physics }}
\rhead{{\bf \color{scipostdeepblue} ~Submission }}

\fancyfoot[C]{\textbf{\thepage}}
}

\begin{document}

\pagestyle{SPstyle}

\begin{center}
{\Large \textbf{\color{scipostdeepblue}{
Surface States in Strain-Induced Nodal-Line Topological Semiconductors
}}}
\end{center}

\begin{center}\textbf{
Vitaly N. Golovach\textsuperscript{1,2,3,4$^\star$} and Alexander Khaetskii\textsuperscript{5$^\dagger$}
}\end{center}

\begin{center}
{\bf 1} Centro de Física de Materiales (CFM-MPC), Centro Mixto CSIC-UPV/EHU, 20018 Donostia-San Sebastián, Spain \\
{\bf 2} Departamento de Polímeros y Materiales Avanzados: Física, Química y Tecnología, Facultad de Química, University of the Basque Country UPV/EHU, 20080 Donostia-San Sebastián, Spain \\
{\bf 3} Donostia International Physics Center (DIPC), 20018 Donostia-San Sebastián, Spain \\
{\bf 4} IKERBASQUE, Basque Foundation for Science, 48013 Bilbao, Spain \\
{\bf 5} Department of Physics and Astronomy, Ohio University, Athens, OH 45701, USA
\\[1em]
$^\star$ \href{mailto:vitaly.golovach@ehu.eus}{vitaly.golovach@ehu.eus}\,, \quad
$^\dagger$ \href{mailto:khaetskii@gmail.com}{khaetskii@gmail.com}
\end{center}

\section*{\color{scipostdeepblue}{Abstract}}
\textbf{\boldmath{%
This work explores the topological phase diagram of inverted-band-gap semiconductors under strain and spin-orbit coupling. 
Using a minimalistic Luttinger Hamiltonian model, we follow the transitions between a 3D topological insulator, 
a Dirac semimetal, a nodal-line semimetal, and a Weyl semimetal. Analytical and exact solutions for surface states are derived for high-symmetry directions as well as in several limiting cases. We demonstrate the continuous evolution of these surface states across phase boundaries, providing a unified picture that synthesizes previous literature. Specifically, we detail the progression from a Dirac to a nodal-line and then to a Weyl semimetal as spin-orbit coupling originating from bulk inversion asymmetry is introduced. A hierarchy of energy scales is established, defining the criteria for realizing these phases. Finally, we reveal a non-analyticity in the surface-state dispersion at the projected nodal line, originating from distinct, terminating patches of surface states with unique spin textures in momentum space.
}}

\vspace{\baselineskip}

\noindent\textcolor{white!90!black}{%
\fbox{\parbox{0.975\linewidth}{%
\textcolor{white!40!black}{\begin{tabular}{lr}
  \begin{minipage}{0.6\textwidth}
    {\small Copyright attribution to authors. \newline
    This work is a submission to SciPost Physics. \newline
    License information to appear upon publication. \newline
    Publication information to appear upon publication.}
  \end{minipage} & \begin{minipage}{0.4\textwidth}
    {\small Received Date \newline Accepted Date \newline Published Date}%
  \end{minipage}
\end{tabular}}
}}
}


\vspace{10pt}
\noindent\rule{\textwidth}{1pt}
\tableofcontents
\noindent\rule{\textwidth}{1pt}
\vspace{10pt}

\section{Introduction}
\label{secIntro}
Topological matter~\cite{KumarFelser2020,GrigoryTkachev}
comprises materials or quantum phases whose defining physical properties are governed by nontrivial global topological invariants~\cite{FuKanePRB2007,HasanKaneRMP2010,QiZhangRMP2011}. 
These materials are of significant interest due to unique features such as bulk-boundary correspondence, robustness against perturbations, 
and quantization~\cite{KaneMele2005,BernevigZhang2006,BHZ2006,BernevigHughesBook,XiaoRMP2010}---properties that hold promise for applications in spintronics and quantum computing. 
Although topological invariants are global quantities, integrated over the entire Brillouin zone, the emergent low-energy physics is frequently dominated by surface states. 
These states encode detailed information about the material’s surface, while their global characteristics---for instance, 
the presence of metallic surface bands and their numbers---are constrained by the underlying topology.

While materials exhibiting pristine topological properties are rare, a substantial class is considered \emph{close to being topological}. 
This classification arises from their possession of a key ingredient: band inversion near the Fermi level. 
Despite the fact that many materials can be formally classified as topological---particularly multi-compound systems where strong spin-orbit coupling promotes band inversion~\cite{Bradlyn2017,Vergniory2019,Vergniory2022}---the 
band inversion is often buried deep below the Fermi energy and has no effect in low-energy physics. 
In this respect, some of the most compelling candidates are those that can be driven into a topological insulating phase by the application of strain.

On the other hand, semiconductors represent one of the most versatile classes of materials, with a tremendous scientific and technological footprint. 
This versatility stems from high-quality growth techniques and the ability to engineer their properties through doping and heterostructuring. 
The creation of heterostructures, in particular, enables the induction and control of significant strain in thin material layers, 
including its magnitude and sign. 
Within this context, HgTe and $\alpha$-Sn stand as the simplest and most-studied prototypes of tunable topological materials, both theoretically~\cite{BHZ2006} 
and experimentally~\cite{RVidal2023}.

In their pristine, semi-metallic state, the low-energy band structure of HgTe and $\alpha$-Sn is most accurately described by the Luttinger Hamiltonian, 
applicable to excitations within the $\Gamma_8$ multiplet. 
For a complete description that includes higher excitation energies, the Kane model is required, 
as it additionally encompasses the $\Gamma_6$ band originating from deep below the Fermi level. 
The buried position of this $\Gamma_6$ band is a direct manifestation of the band inversion that characterizes these close-to-topological materials. 
These two models are complementary, providing a powerful analytical framework for studying surface states in semiconductor heterostructures, particularly under strain.

The application of tensile in-plane strain---which, for example, arises naturally during the growth of CdTe/HgTe heterostructures---opens a band gap in the topological semiconductor, 
transforming it into a three-dimensional topological insulator (TI). 
The surface states that form at a $(001)$ crystallographic interface undergo a continuous evolution from Dyakonov-Khaetskii (DK)~\cite{DK1981,KGK2022} states at zero gap 
to Volkov-Pankratov (VP)~\cite{VP1985} states at extremely large strain-induced gaps~\cite{DaiZhang2008}. 
This continuous transition demonstrates that, despite the DK states being derived within the Luttinger Hamiltonian model 
and the VP states within the Kane model, they represent a unified set of surface states originating from the material's topological character. 
Crucially, the presence of the heavy-hole band leads to a non-trivial interaction with the surface states~\cite{KGK2022,PPV1987}.

This interaction fragments the surface states into several subbands, giving rise in particular to \emph{wing states}  within the projected band gap~\cite{KGK2024}. 
These wing states initially accompany the DK states at small excitation energies in a manner reminiscent of conventional spin-orbit splitting. 
However, while their presence does not violate topological constraints---they begin and end in the same bulk band---they demonstrate that 
the surface states of a TI can exhibit characteristics akin to two-dimensional systems with spin-splitting induced by structural inversion asymmetry. 
In this context, the VP states are also often described qualitatively as linearly-dispersing two-dimensional surface states arising from a large, 
effective spin-orbit interaction~\cite{VP1985}.

The formation of paired, spin-split surface states becomes more pronounced under compressive in-plane strain, 
as engineered in ZnTe/HgTe or CdTe/$\alpha$-Sn heterostructures. 
In this regime, the topological semiconductor transitions into a three-dimensional Dirac semimetal~\cite{Young2012,ArmitageRMP2018,Ruan2016,BansilRMP2016,Lv2021}, 
characterized by two conical bulk dispersion points near the $\Gamma$-point. 
Remarkably, the spectrum of the DK surface states persists. 
The conical bulk dispersion creates a sufficiently wide projected band gap, allowing the observation of both branches of the DK states~\cite{KGK2022}. 
Both DK branches take off from the Dirac point in the projected band gap.
Consequently, the surface states within the gap appear as a pair of two-dimensional states with an effective spin-splitting that scales quadratically with momentum, see Fig.~7 of Ref.~\cite{KGK2022}.
This scaling is unconventional for a pure spin-$1/2$ subspace, where time-reversal symmetry typically restricts spin-orbit terms to odd powers of momentum. 
However, the strong locking between the wave vector and the angular momentum $J=3/2$, inherent to the Luttinger Hamiltonian, involves a more complex pseudospin degree of freedom. 
The resulting transformation law for the two DK branches yields an effective surface Hamiltonian that hosts this distinctive quadratic splitting 
while still preserving overall time-reversal symmetry.

The situation becomes even more interesting when bulk inversion asymmetry is considered. 
While $\alpha$-Sn crystallizes in a diamond lattice with $O_h$ point-group symmetry---yielding a 
$C_{4v}$ symmetry at the $(001)$ surface---HgTe possesses a zinc-blende lattice with $T_d$ symmetry.
 For HgTe, this lowers the surface point-group symmetry to $C_{2v}$. 
 This reduced symmetry of the surface potential is not inherently captured by the standard Luttinger Hamiltonian. 
 The most straightforward method to incorporate the effects of the 
$T_d$ crystal class into this model is to include Dresselhaus spin-orbit terms for the $J=3/2$ manifold.
Upon including linear-in-momentum  Dresselhaus terms, 
compressively strained HgTe undergoes a topological transition from a three-dimensional Dirac semimetal to a nodal-line semimetal~\cite{Ruan2016,Fang2015,Fang2016}. 
Specifically, each original Dirac point spreads into a circular nodal line.

In this work, we demonstrate that the Luttinger Hamiltonian, augmented with Dresselhaus spin-orbit terms, 
admits an exact analytical solution for wave-vector directions along the high-symmetry axes of the $(001)$ surface. 
The linear Dresselhaus terms deform and shift the two DK surface state branches in momentum space. 
Crucially, the bulk nodal line remains a singular point for these surface states, where one DK branch connects non-analytically to the other. 
This non-analyticity at the surface---directly inherited from the bulk topology---induces distinct non-analytic properties in surface observables 
as a function of the in-plane wave vector.
Consequently, while the surface state spectrum resembles two spin-split branches of a conventional two-dimensional system, 
the geometric singularity enforced by the bulk nodal line renders it topologically distinct. 
This manifests as a patch structure in momentum space, where the surface states are organized into separate regions. 
We show that the topological constraint of patching these regions around the nodal line leads to the formation 
of three two-dimensional Dirac points along one of the high-symmetry directions on the surface.

\section{The hierarchy of energy scales at the nodal line}
\label{secHierarchy}

\begin{figure}[!t]
\centering
\includegraphics[width=0.46\textwidth]{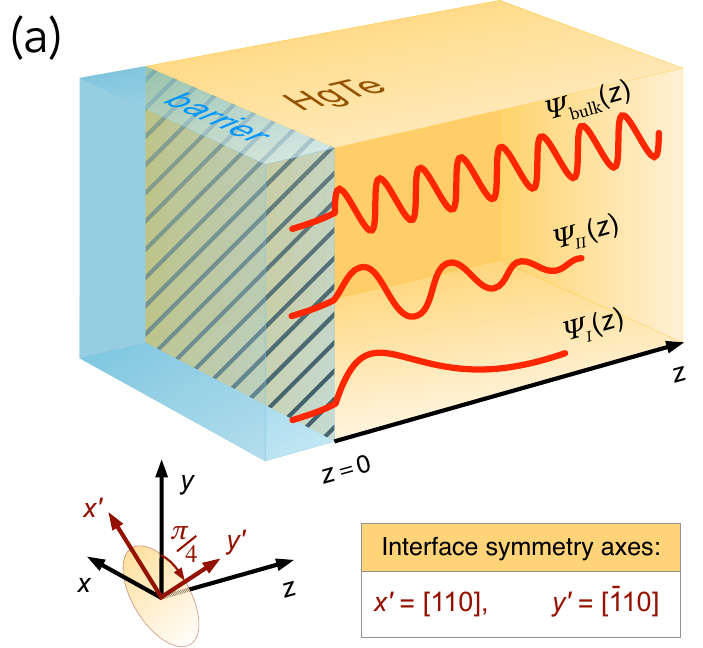}
\hfill
\includegraphics[width=0.48\textwidth]{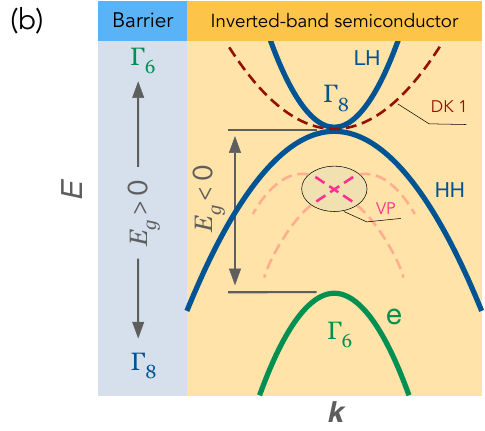}
\caption{(a) Interface geometry: an HgTe crystal capped by a barrier or exposed to vacuum at the surface $z=0$. 
The lowering of point-group symmetry from $C_{4v}$ to $C_{2v}$ at the interface (hatched region, indicating symmetry change) is due to the $T_d$ crystal class of HgTe. 
This symmetry reduction entails a $\pi/4$ rotation of the coordinate system about the $z$-axis. 
Various states can exist at the interface: bulk standing waves $\Psi_{\rm bulk}(z)$, simple evanescent surface states $\Psi_{\rm I}(z)$, and oscillatory evanescent surface states $\Psi_{\rm II}(z)$.
(b) Schematic energy diagram for the interface in (a), depicting the inverted band structure of bulk HgTe (solid lines). 
The $\Gamma_6$ band lies deep below the chemical potential, while the unoccupied light-hole (LH) branch of the $\Gamma_8$ multiplet acts as the conduction band, 
forming a gapless semiconductor (semimetal) with the occupied heavy-hole (HH) branch. 
The surface states (dashed lines) consist of the Dyakonov-Khaetskii branch of the first type (DK 1), which forms in the projected band gap between the HH and LH bands, 
and Volkov-Pankratov (VP) branches. 
The VP states hybridize strongly with the bulk HH band at large wavevectors but appear as a sharp surface resonance near zero in-plane wavevector, 
where they decouple from the HH states. 
The fine structure related to strain and further point-group symmetry lowering at the interface is shown in Fig.~\ref{fig:strain}.
\label{fig:interface}}
\end{figure}

We consider the $(001)$ crystallographic surface of an inverted-band semiconductor, see Fig.~\ref{fig:interface}(a).
In the absence of strain and spin-orbit coupling, the band structure of the semiconductor is shown in Fig.~\ref{fig:interface}(b).
Notably, since the $\Gamma_6$ and $\Gamma_8$ bands are reversed in energy with respect to the usual order in semiconductors, 
the light-hole (LH) band of the $\Gamma_8$ multiplet plays the role of the conduction band,
whereas the $\Gamma_6$ band resides deep under the Fermi level $\mu$. 
The surface states form in the projected band-gap between the heavy-hole (HH) and the LH bands and are given, for realistic parameters, 
by the DK branch of the first type (DK 1)~\cite{DK1981,KGK2022}.
In contrast, the VP states~\cite{VP1985} hybridize with the HH band and hence appear as surface resonances in the vicinity of zero in-plane wave-vector.
At this special point, the VP states decouple from the HH band, whereas the DK states become strongly delocalized. 
In this respect, one can regard the DK and VP states, in a unified description, as being two parts of a topological surface state which is interrupted by the HH band~\cite{KGK2022,KGK2024}.

\subsection{Model for topological surface states}
\label{secModels}

While the interaction of topological surface states with the HH band can be captured by the Kane model, 
our focus here is on the low-energy physics of the $\Gamma_8$ multiplet, which is most accurately described by the Luttinger Hamiltonian~\cite{Winkler}
\begin{equation}
H_{\text{L}}=\left(\gamma_1+\frac{5}{2}\gamma_2\right)k^2-2\gamma_2\left(k_x^2J_x^2+\text{c.p.}
\right)
-4\gamma_3\left(
k_xk_y\left\{J_xJ_y\right\}+\text{c.p.}
\right),
\label{eqHamLutt}
\end{equation}
where $k^2=k_x^2+k_y^2+k_z^2$ is the square of the wave vector,
$x$, $y$, and $z$ are coordinate axes aligned with the principal crystallographic directions of the cubic crystal,
$J_x$, $J_y$, and $J_z$ are the spin-$3/2$ matrices,  c.p.\@ denotes cyclic permutations,  
$\left\{AB\right\}=\frac{1}{2}\left(AB+BA\right)$ is the symmetrized product,
and $\gamma_1$, $\gamma_2$, and $\gamma_3$ 
are the Luttinger parameters.
For HgTe, these parameters are $\gamma_1=15.6\times\hbar^2/2m_e$, $\gamma_2=9.6\times\hbar^2/2m_e$, and 
$\gamma_3=8.6\times\hbar^2/2m_e$, where $m_e$ is the free electron mass.

\begin{figure*}[!ht]
\centering
\includegraphics[width=\textwidth]{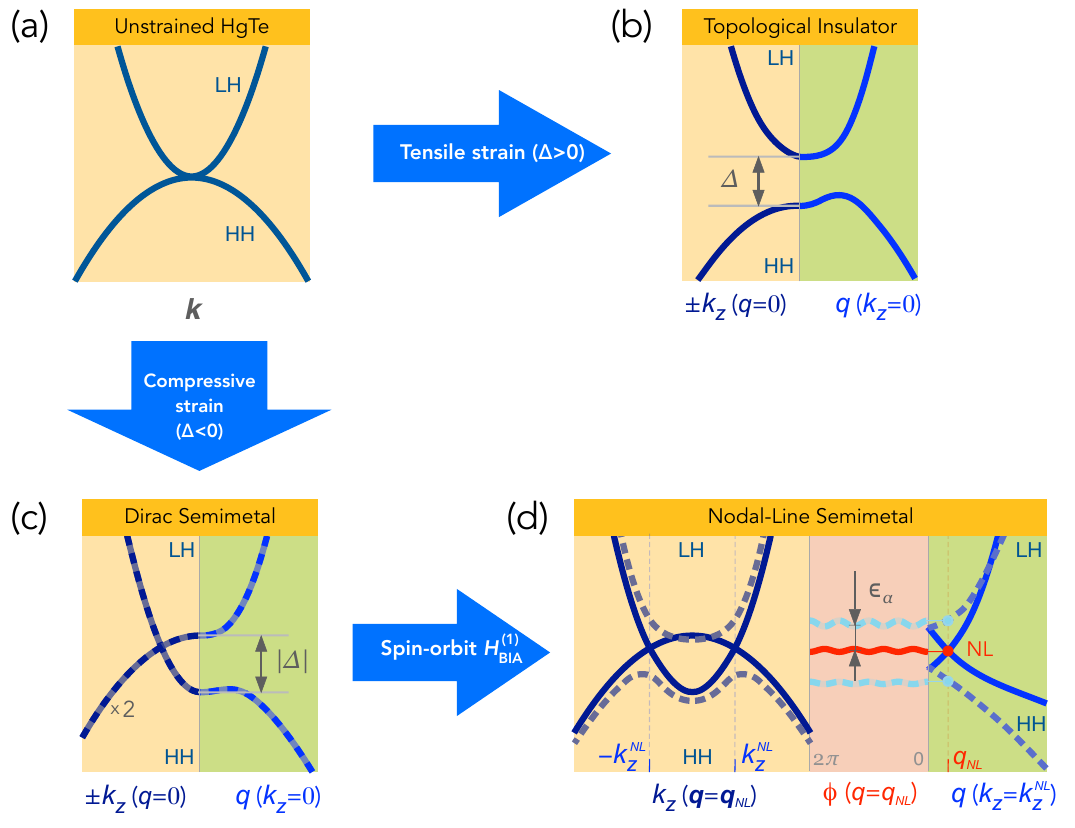}
\caption{Effect of strain and spin-orbit splitting on the $\Gamma_8$ bands in HgTe: (a) Unstrained light-hole (LH) and heavy-hole (HH) bands.
(b) A tensile strain ($\Delta>0$) mixes the bands and leads to the formation of an absolute gap (equal to a fraction of $\Delta$) and transforms HgTe into a TI.
(c) A compressive strain ($\Delta<0$) leads to band overlaps and transforms HgTe into a 3D Dirac semimetal with two Dirac points occurring at $k_z=\pm k_z^{\rm NL}$ and zero in-plane momentum ($q=0$).
(d) The two-fold spin degeneracy of the bulk bands is lifted by the linear-in-momentum Dresselhaus spin-orbit coupling in Eq.~(\ref{eqHBIAalph}), 
which splits away the bulk bands denoted by thick dashed lines by an energy $\epsilon_\alpha$ given in Eq.~(\ref{eqepsalph}).
Each Dirac point turns into a nodal line (NL), which is formed by the two remaining intersecting bands (solid lines), 
but which intersect each other now at a finite in-plane momentum $q_{\rm NL}\approx\alpha/2\gamma_3$.
\label{fig:strain}}
\end{figure*}

The evolution of the $\Gamma_8$ multiplet under strain and spin-orbit coupling is depicted in Fig.~\ref{fig:strain}. 
Under tensile strain, the LH and HH bands shift apart in energy at zero in-plane momentum ($\bm{q}=0$). 
At finite $\bm{q}=\left(k_x,k_y\right)$, however, they intermix and experience mutual attraction, leading to a mass inversion for the HH band near $q=0$. 
Crucially, the tensile strain opens an absolute gap in the bulk spectrum [Fig.~\ref{fig:strain}(b)], 
driving the transition from a gapless semiconductor to a TI. 
The surface state bridging this gap is the first Dyakonov-Khaetskii (DK 1) branch, as detailed in Ref.~\cite{KGK2022}.

In contrast, under compressive strain, the LH and HH bands intermix strongly, but no absolute band gap opens. 
Instead, two three-dimensional Dirac points emerge in the bulk spectrum at a finite value of $|k_z|$ [Fig.~\ref{fig:strain}(c)]. 
The subsequent inclusion of spin-orbit interaction lifts the twofold spin degeneracy of the bulk bands, 
transforming the two Dirac points into two nearly circular nodal lines [Fig.~\ref{fig:strain}(d)].

We incorporate the effects of strain via the Bir-Pikus Hamiltonian~\cite{BirPikus}. 
This Hamiltonian is constructed from the same invariants as the Luttinger Hamiltonian in Eq.~\eqref{eqHamLutt},
but with the momentum products  $k_ik_j$
replaced by the corresponding components of the strain tensor $\varepsilon_{ij}$.
For compressive in-plane strain, the only non-zero strain components are
$\varepsilon_{xx}=\varepsilon_{yy}<0$ and $\varepsilon_{zz}>0$,
reducing the Bir-Pikus Hamiltonian to the form
\begin{equation}
H_{\textrm{strain}}=-\frac{\Delta}{2}\left(J_z^2-\frac{5}{4}\right),
\end{equation}
where $\Delta=-2b\left(\varepsilon_{xx}-\varepsilon_{zz}\right)$ and  $b=-\frac{2}{3}D_{u}$, with $D_{u}$ being a deformation potential for 
holes~\cite{Winkler}.
For HgTe, $b\approx -1.5\;\textrm{eV}$, resulting in $\Delta<0$ for the compressive strain considered in this work. 
Typical magnitudes of $\Delta$ are on the order of $10\;\textrm{meV}$.

Bulk inversion asymmetry (BIA) in a cubic crystal with zinc-blende structure ($T_d$) 
splits the band structure and lifts the two-fold spin degeneracy of the bulk bands via a spin-orbit interaction mechanism~\cite{Dresselhaus1955,Cardona1986,Cardona1988}. 
To first order in $k_i$,  this splitting is described by an additional term in the Hamiltonian~\cite{Winkler},
\begin{equation}
H_{\text{BIA}}^{(1)}=2\alpha\left[k_x\left\{J_x\left(J_y^2-J_z^2\right)\right\}+\text{c.p.}\right],
\label{eqHBIAalph}
\end{equation}
where $\alpha=C_k/\sqrt{3}$.
The parameter $C_k$ originates from an extended Kane model and accounts for second-order transitions to remote bands outside of the nearest 16-band manifold~\cite{Cardona1988,Winkler}.
Reported values for HgTe vary: the value $C_k\approx -74.6\,\textrm{meV\,\AA}$ 
is obtained in~\cite{Cardona1986,Cardona1988}, 
whereas more recent works~\cite{Ruan2016,Kibis}  
use  the value $\alpha\approx 0.208\,\textrm{eV\,\AA}$, corresponding to a $C_k$ value approximately five times larger and of opposite sign.
Strictly speaking, the sign of  $C_k$  is fixed by a convention based on the physical distinction between the
$[110]$ and $[1\bar{1}0]$ directions on a (001) crystal surface~\cite{Cardona1986}.
Consequently, the dispersion of surface states along the in-plane 
$[110]$ direction is expected to differ from that along the 
$[1\bar{1}0]$ direction---a result we confirm in Sec.~\ref{secPhipi4}.
However, since the sign change of $\alpha$ is equivalent to a $\pi/2$-rotation of the coordinate frame, we assume further $\alpha>0$.

\subsection{Nodal-line physics}
\label{secNodalLinePhys}

The nodal line occurs at $q\approx \alpha/2\gamma_3$ and has a nearly circular shape for weak spin-orbit coupling ($\alpha\ll\sqrt{\gamma_2\left|\Delta\right|}$).
Exact relations for the nodal line can be obtained for the spherical approximation ($\gamma_3=\gamma_2$).
We find
\begin{equation}
\begin{aligned}
q_{\textit{NL}} &= \frac{\alpha}{2\gamma_2},\\
\left(k_z^{\textit{NL}}\right)^2 &=\frac{q_{\textit{NL}}^2}{2}-\frac{\Delta}{4\gamma_2}+\delta\left(\phi\right),\\
E_{\textit{NL}} &= \frac{3}{2}\left(\gamma_1q_{\textit{NL}}+\alpha\right)q_{\textit{NL}}-\frac{\gamma_1 \Delta}{4\gamma_2}-(2\gamma_2-\gamma_1)\delta\left(\phi\right),
\end{aligned}
\label{eqNLdefs}
\end{equation}
where $\phi$ is the azimuthal angle of the in-plane wave vector $\bm{q}$, defined by $k_x=q\cos\phi$ and $k_y=q\sin\phi$, and 
\begin{equation}
\delta\left(\phi\right) = \frac{\Delta}{12\gamma_2}+\sqrt{\left(\frac{\Delta}{12\gamma_2}\right)^2+\frac{1}{4}q_{\textit{NL}}^4\cos^22\phi}.
\label{eqNLdefsdt}
\end{equation}
Recall that $\Delta<0$.  In the limit of small spin-orbit interaction, $\delta\left(\phi\right)\propto\alpha^4$. 

The energy window within which the bands intersecting at the nodal line are the only states at the Fermi level---often termed the adiabaticity window for nodal-line physics---is given by
\begin{equation}
\epsilon_{\alpha} \approx\alpha\sqrt{\frac{3\left|\Delta\right|}{\gamma_2}}.
\label{eqepsalph}
\end{equation}
Recall that $\left|\Delta\right|$ sets the adiabaticity window for 3D Dirac semimetal, see Fig.~\ref{fig:strain}.
Therefore, an energy scale separation naturally occurs for HgTe, where the physics related to the Dirac semimetal can be
observed on the scale of $\left|\Delta\right|\simeq 10\,\textrm{meV}$, 
whereas nodal-line physics requires an energy resolution of $\epsilon_{\alpha} \simeq 1\,\textrm{meV}$.

To complete the hierarchy of adiabaticity scales for compressively strained HgTe, we also include one of the cubic-in-momentum Dresselhaus terms,
\begin{equation}
H_{\text{BIA}}^{(3)}=\beta
\left[k_x\left(k_y^2-k_z^2\right)J_x+\text{c.p.}\right] ,
\label{eqHBIAbeta}
\end{equation}
where $\beta\equiv b_{41}^{8v8v}$~\cite{Winkler}.
This coupling can be derived from third-order perturbation theory 
within the extended Kane model and therefore involves transitions to bands that are energetically closer than those responsible for the coupling in Eq.~\eqref{eqHBIAalph}.
For CdTe, the value of $b_{41}^{8v8v}\approx -76.93\,\textrm{eV}\,\textrm{\AA}^3$ is listed in~\cite{Winkler}.
However, we were unable to locate a value for HgTe---either measured or calculated---in the literature.

It is known that the spin-orbit interaction in Eq.~\eqref{eqHBIAbeta} lifts the degeneracy of the bulk states at the nodal line, 
transforming the nodal-line semimetal into a Weyl semimetal~\cite{Ruan2016}. 
Consequently, the band degeneracy along each of the two nodal lines of HgTe is lifted everywhere except at four points corresponding to $\phi = 0, \pi/2, \pi, 3\pi/2$. 
The degeneracy at these points is ``protected'' by a hidden symmetry of the Hamiltonian that emerges for special values of $q$ and $\pm k_z$; 
these values are generally shifted from the original nodal-line coordinates by displacements proportional to $\beta$. 
As a result, four Weyl monopoles emerge from each nodal line~\cite{Ruan2016}.

The adiabaticity scale associated with the Weyl semimetal phase can be determined by calculating 
the splitting of the nodal line along directions that pass between neighboring monopoles. 
These directions, which correspond to $\phi = \pm\pi/4, \pm 3\pi/4$, 
coincide with the high-symmetry axes of the surface crystallographic potential.
To leading order in $\beta$, the half-splitting is
\begin{equation}
\epsilon_{\beta}=\frac{\alpha\beta\Delta}{16\gamma_2\gamma_3}
\sqrt{2+\frac{1}{1-\frac{3\alpha^2\gamma_2}{4\gamma_3^2\Delta}}}\approx
\frac{\sqrt{3}\alpha\beta\Delta}{16\gamma_2\gamma_3}.
\end{equation}
Assuming the value of $\beta$ for HgTe is similar to that for CdTe, we obtain $\epsilon_{\beta}\approx 3\,\mu\textrm{eV}$.
This energy scale is extremely small, suggesting that the distinct signatures of the Weyl semimetal phase 
may be challenging to resolve experimentally in HgTe. 
Nevertheless, the emergent hierarchy of energy scales in this material is instructive. 
It demonstrates that
$\alpha/\sqrt{\gamma_2\left|\Delta\right|}$  and $\left(\beta/\gamma_3\right)\sqrt{\left|\Delta\right|/\gamma_2}$
can be formally treated as small parameters governing the sequential transitions from a Dirac point 
to a nodal line and subsequently to Weyl monopoles.

We obtain exact analytical solutions for the surface states of the model
\begin{equation}
H=H_{\text{L}}+H_{\textrm{strain}}+H_{\text{BIA}}^{(1)}+H_{\text{BIA}}^{(3)},
\end{equation}
along the high-symmetry directions of the in-plane momentum.
Selected results are presented in Sec.~\ref{secPhipi4}, where $H_{\text{BIA}}^{(3)}$ is omitted for the simplicity of presentation.
Our central finding is the non-analytic behavior of the surface states at the nodal line.
In the absence of $H_{\text{BIA}}^{(3)}$,
this manifests as a discontinuous derivative 
$\partial E/\partial q$ 
of the surface-state branch crossing the nodal line, as well as in the behavior of momentum-resolved physical quantities.

A natural question arises: How does the energy scale $\epsilon_\beta$
affect the surface states?
Na\"{i}vely, one would expect the avoided crossing of the bulk bands at the nodal line to create a crossover region for the surface states, 
where the kink of $E(q)$ (\emph{i.e.}, the non-analytic behavior) at $q=\alpha/2\gamma_3$
 is smoothed out. Surprisingly, we find that not only does the non-analytic behavior persist after lifting the nodal-line degeneracy, but it is in some sense amplified. 
 Instead of a discontinuous derivative, the energy dispersion 
 $E(q)$
 itself becomes discontinuous at the former nodal line position, with an energy jump of approximately
 $\Delta E(q_{\textit{NL}}) \approx 2\epsilon_\beta$.

Away from the nodal line, the surface states are affected only insignificantly by the cubic terms. 
Thus, for most purposes---especially when studying physical properties that do not explicitly probe the splitting of 
the nodal line---one may draw little practical distinction between the nodal-line semimetal and Weyl semimetal phases. 
For this reason, we omit $H_{\text{BIA}}^{(3)}$ in subsequent sections.

\begin{figure*}[!b]
\centering
\includegraphics[width=0.48\textwidth]{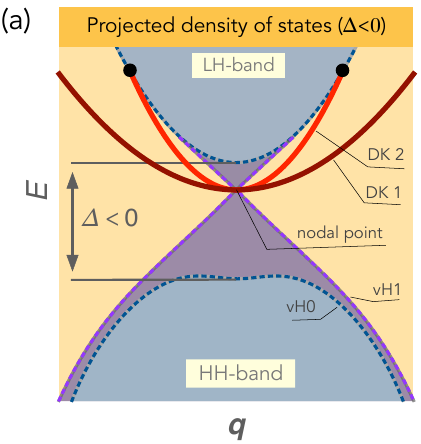}
\hfill
\includegraphics[width=0.48\textwidth]{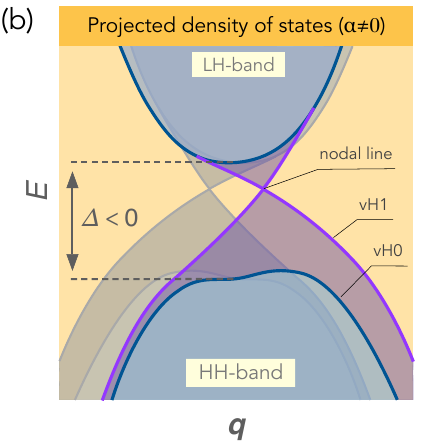}
\caption{(a) Projected density of states (PDOS) and surface states for $\Delta<0$ in the absence of Dresselhaus spin-orbit interaction. 
 The PDOS is bounded by two types of van Hove singularities, labeled vH0 and vH1. 
 The singularity vH0 corresponds to the bulk spectrum at $k_z=0$, while vH1 forms an envelope function 
 for the extrema of the bulk spectrum at finite $k_z\neq 0$. 
 These linearly dispersing vH1 singularities converge to form a nodal point at $q=0$. 
 Two branches of Dyakonov-Khaetskii surface states, labeled ```DK 1'' and ``DK 2'', 
 emerge from this nodal point and disperse upward in energy~\cite{KGK2022}. 
 (b)  Projected density of states upon inclusion of the linear spin-orbit terms from Eq.~(\ref{eqHBIAalph}). 
 The PDOS splits in a complex manner. 
 For clarity, one spin component is drawn with thicker, brighter lines, while the other is shown in gray tones. 
 The corresponding surface states are not displayed here and are presented separately in Fig.~\ref{SurfaceStatesExact0}.
\label{fig:ProjDOSSemiMetal}}
\end{figure*}

\subsection{Two types of van Hove singularities in the projected density of states}
\label{secVanHove}

We begin by recalling that under compressive strain and in the absence of spin-orbit interaction, the surface states consist of two quadratically dispersing branches (DK 1 and DK 2), as shown in Fig.~\ref{fig:ProjDOSSemiMetal}(a). 
The projected density of states (PDOS) exhibits two distinct types of van Hove singularities.
One type of singularity, vH0, coincides with the energy of the bulk spectrum evaluated at
$k_z= 0$.
Within the spherical approximation 
($\gamma_3=\gamma_2$), this is given by
\begin{equation}
E_{\pm}^{\text{vH0}}(q)=
\gamma_1q^2\pm \gamma_2\sqrt{3q^4+4k_\Delta^4},
\label{eqEvH0}
\end{equation}
where
\begin{equation}
k_\Delta=\sqrt{\frac{q^2}{2}-\frac{\Delta}{4\gamma_2}}.
\label{eqkDeltaAppx}
\end{equation}

Another type of singularity, vH1, originates from band repulsion near $k_z\simeq \pm k_{\Delta}$.
It forms an envelope function for the extrema of the bulk spectrum at different $k_z$,
\begin{equation}
E_{\pm}^{\text{vH1}}(q)=-\frac{\gamma_1\Delta}{4\gamma_2}
\pm\sqrt{3}\gamma_2\eta_+\eta_- q 
\sqrt{4k_\Delta^2-2q^2},
\label{eqEvH1}
\end{equation}
where 
\begin{equation}
\eta_\pm=\sqrt{1 \pm \frac{\gamma_1}{2 \gamma_2}}.
\end{equation}
This van Hove singularity exists only if
\begin{equation}
\gamma_1 E_{\pm}^{\text{vH1}}(q) <-\gamma_2\Delta
+\left(\gamma_1^2-4\gamma_2^2\right)q^2.
\end{equation}
If vH1 exists, it determines the band edge of the PDOS; 
otherwise, vH0 serves this role. 
For HgTe, the material parameters ensure that vH1 exists without restriction for the HH band, 
whereas for the LH band it exists only within a finite range of  $q$.

We note that the van Hove singularity vH0
in the HH band (see Fig.~\ref{fig:ProjDOSSemiMetal}) has a saddle-point shape.
While this feature could be mistaken for the surface resonance originating from the 
VP states [illustrated in Fig.~\ref{fig:interface}(b)], a key distinction exists.
The VP resonance occurs significantly deeper within the inverted band gap 
$E_g$, and its energy position is controlled by the ratio of band offsets at the interface, 
as detailed in~\cite{KGK2024}.

\section{Surface states along high-symmetry directions in momentum space}
\label{secPhipi4}
We consider the high-symmetry directions $[11]$ and $[1\bar{1}]$ of the in-plane momentum $\bm{q}$,
corresponding in our notations to $\phi=\pi/4$ and $\phi=-\pi/4$, respectively.
For these directions, 
the surface-state wave function can be cast
into a product of a spinor $\chi$ and an orbital wave function
\begin{equation} 
\psi(x,y,z)= \chi(k_x,k_y)\,
\psi_{\text{orb}}(z)e^{i\left(k_xx+k_yy\right)},
\label{eqPsiChi}
\end{equation}   
where the orbital $z$-component for $z>0$ is
\begin{equation} 
 \psi_{\text{orb}}(z)
 = \frac{1}{\Gamma} \sqrt{\kappa (\kappa^2 - \Gamma^2)}
  e^{-\kappa z}
  \left(e^{\Gamma z} - e^{-\Gamma z}\right).
\label{eqpsi}
\end{equation}
This ansatz automatically satisfies the boundary condition $ \psi_{\text{orb}}(z)=0$ at the interface $z=0$.
The parameter $\kappa$ is real and positive, governing the overall evanescent decay of the wave function.
The parameter $\Gamma$ can be 
either real and smaller than $\kappa$ ($0<\Gamma < \kappa$) or purely imaginary ($\Gamma=i\gamma$).
In the latter case, 
the wave function describes a standing wave with an evanescent envelope,
corresponding to the oscillatory surface state $\Psi_{\text{II}}(z)$ in Fig.~\ref{fig:interface}(a).
We note that the separation of variables into a spinor and an orbital component is enabled by the
$C_{2v}$ point-group of the surface potential, which possesses mirror planes containing 
the $[11]$ and $[1\bar{1}]$  axes.

\subsection{Exact surface-state solutions}
\label{secExactSols}

We present our results for $E$, $\chi$, $\kappa$, and $\Gamma$ 
obtained assuming 
$\alpha>0$, $\gamma_2>0$, $\gamma_3>0$, $\left|\gamma_1\right|<2\gamma_2$, and $\Delta<0$.
In order to simplify the writing, we use the notation
\begin{equation}
s=\left\{
\begin{array}{cc}
1,\quad\mbox{for}\quad & \phi=\pi/4, -3\pi/4\\
-1,\quad\mbox{for}\quad & \phi =-\pi/4, 3\pi/4
\end{array}\right.
\label{eqetapmdef}
\end{equation}
The surface states consist of two branches of energy, 
which form a 2D Dirac point at $q=0$~\cite{Kibis}.
We follow these energy branches away from the point $q=0$ in the four wave-vector directions in Eq.~(\ref{eqetapmdef}), 
denoting the branch which doesn't pass through the nodal line by $E_1$, whereas the branch that passes through the nodal point by $E_2$.
For $E_2$, we distinguish between the inner part, for $q<\alpha/2\gamma_3$, and the outer part, for $q>\alpha/2\gamma_3$.
We use a bar over the index $2$ to denote quantities in the outer region.

For the analytic branch, we obtain
\begin{equation}
E_{1}(q)=- \frac{\gamma_1 \Delta}{4\gamma_2}+
\frac{3}{2} (\gamma_1 q-\alpha)q -
  \frac{\sqrt{3}}{2}
   \left(2\gamma_3 q+\alpha\right)s\eta_+\eta_-q,
\label{eqEDK1}
\end{equation}
and the spinor 
\begin{equation}
\begin{split}
\chi_{1}&=\frac{\eta_+}{\sqrt{2}}\left(\left|\frac{3}{2},+\frac{3}{2}\right\rangle+\frac{k_x+ik_y}{q}\left|\frac{3}{2},-\frac{3}{2}\right\rangle\right) 
            - \frac{i \eta_-}{\sqrt{2}}\left(\frac{k_x+ik_y}{q} \left|\frac{1}{2},+\frac{1}{2}\right\rangle-\left|\frac{1}{2},-\frac{1}{2}\right\rangle\right).
\end{split}
\label{eqChiDK1}
\end{equation}
The exponential decay is governed by
\begin{equation}
\kappa_{1}=\frac{
\sqrt{3} }{
  4\gamma_2\eta_+\eta_-}(\alpha+2 \gamma_3 q),
\label{eqkappaDK1}
\end{equation}
and the oscillatory behavior by 
\begin{equation}
\Gamma_{1}= 
  \sqrt{
  \kappa_{1}^2
  - \frac{s\gamma_1}{2 \gamma_2}\kappa_{1}q  
- \frac{q^2}{2} + \frac{\Delta}{4 \gamma_2}
   }.
\label{eqGammaDK1}
\end{equation}
Since $E_1(q)$ does not cross the nodal line, these expressions are valid without restrictions for $0<q<q_*$, where 
$q_*$, if it exists, is a termination point defined by $\kappa_1 =\Gamma_1$.
For $s=1$, this branch has no termination point in this model.
 
For the non-analytic branch, we find for $q<\alpha/2\gamma_3$
\begin{equation}
E_{2}(q)=- \frac{\gamma_1 \Delta}{4\gamma_2}+
\frac{3}{2} (\gamma_1 q+\alpha)q -
  \frac{\sqrt{3}}{2}
   \left(2\gamma_3 q-\alpha\right)s\eta_+\eta_-q,
\label{eqEDK2}
\end{equation}
and its corresponding spinor 
\begin{equation}
\begin{split}
\chi_{2}&=\frac{\eta_+}{\sqrt{2}}\left(\left|\frac{3}{2},+\frac{3}{2}\right\rangle-\frac{k_x+ik_y}{q}\left|\frac{3}{2},-\frac{3}{2}\right\rangle\right) 
            + \frac{i \eta_-}{\sqrt{2}}\left(\frac{k_x+ik_y}{q} \left|\frac{1}{2},+\frac{1}{2}\right\rangle+\left|\frac{1}{2},-\frac{1}{2}\right\rangle\right).
\end{split}
\label{eqChiDK2}
\end{equation}
Similarly, the expressions for $\kappa_2$ and $\Gamma_2$ read
\begin{equation}
\kappa_{2}=\frac{
\sqrt{3} }{
  4\gamma_2\eta_+\eta_-}(\alpha-2 \gamma_3 q),
\label{eqkappaDK2}
\end{equation}
\begin{equation}
\Gamma_{2}= 
  \sqrt{
   \kappa_{2}^2
   +  \frac{s\gamma_1}{2 \gamma_2}\kappa_{2}q 
- \frac{q^2}{2} + \frac{\Delta}{4 \gamma_2}
   }.
\label{eqGammaDK2}
\end{equation}
We note that these expressions are related to those in Eqs.~(\ref{eqEDK1})-(\ref{eqGammaDK1}) by the formal substitution $|\bm{q}| \to -|\bm{q}|$. 
This is not a reversal of the in-plane momentum but rather a formal sign reversal of the square root 
within the definition $q = \sqrt{k_x^2 + k_y^2}$.
For this reason, this branch of surface states can be interpreted as a shifted $E_1(q)$ branch extending from the opposite direction of the in-plane momentum.
The shift is qualitatively similar to a displacement of spectrum caused by linear-in-momentum spin-orbit interaction in conventional two-dimensional systems, unrelated to TIs.
However, the mapping via the formal substitution $q\to -q$ 
is valid only for the two surface state branches within the inner region bounded by the nodal line ($q<\alpha/2\gamma_3$) and fails in the outer region.

Exactly at the nodal line ($q=\alpha/2\gamma_3$),  no localized surface state exists. 
This is consistent with the fact that the gap in the bulk projected density of states vanishes at this line, precluding the formation of any in-gap bound state.

In the outer region ($q>\alpha/2\gamma_3$), for the non-analytic branch,
we find
\begin{equation}
E_{\bar{2}}=
- \frac{\gamma_1 \Delta}{4\gamma_2}+
\frac{3}{2} ( \gamma_1 q+\alpha)q +
  \frac{\sqrt{3}}{2}
   ( 2 \gamma_3 q-\alpha) s\eta_+\eta_- q,
\label{eqEDK2b}
\end{equation}
and the spinor 
\begin{equation}
\begin{split}
\chi_{\bar{2}}&=\frac{\eta_+}{\sqrt{2}}\left(\left|\frac{3}{2},+\frac{3}{2}\right\rangle-\frac{k_x+ik_y}{q}\left|\frac{3}{2},-\frac{3}{2}\right\rangle\right) 
            - \frac{i \eta_-}{\sqrt{2}}\left(\frac{k_x+ik_y}{q} \left|\frac{1}{2},+\frac{1}{2}\right\rangle+\left|\frac{1}{2},-\frac{1}{2}\right\rangle\right).
\end{split}
\label{eqChiDK2b}
\end{equation}
Furthermore, 
\begin{equation}
\kappa_{\bar{2}}=\frac{
\sqrt{3}}{
  4\gamma_2\eta_+\eta_-}\left(2 \gamma_3 q-\alpha\right),
\label{eqkappaDK2b}
\end{equation}
\begin{equation}
\Gamma_{\bar{2}}= 
  \sqrt{
  \kappa_{\bar{2}}^2
   + \frac{s\gamma_1}{2 \gamma_2}\kappa_{\bar{2}}q
- \frac{q^2}{2} + \frac{\Delta}{4 \gamma_2}
   },
\label{eqGammaDK2b}
\end{equation} 
We remark that the expressions in Eqs.~(\ref{eqEDK2b})-(\ref{eqGammaDK2b}) could be obtained from Eqs.~(\ref{eqEDK1})-(\ref{eqGammaDK1}) 
by replacing simultaneously $q\to-q$ and $\eta_-\to-\eta_-$.
Precisely this latter sign change brings about the non-analytic behaviour in the quantities 
characterising the surface state.
Notably, the spinor changes abruptly upon traversing the nodal line, see Eqs.~(\ref{eqChiDK2}) and~(\ref{eqChiDK2b}).

\subsection{Non-analytic properties of solutions at the nodal line}
\label{secNonAnalyticity}

To quantify how abruptly the spinor changes, we calculate 
\begin{equation}
1-\left|\left\langle\chi_{2}| \chi_{\bar{2}}\right\rangle\right|^2=1-\frac{\gamma_1^2}{4\gamma_2^2}.
\label{eqdisconmeasure}
\end{equation}
Consequently, the non-analytic behavior at the nodal line should be most pronounced in materials with a small ratio
$\gamma_1/2\gamma_2$.
Typically, this ratio is close to unity (\emph{e.g.}\@ $\gamma_1/2\gamma_2\approx 0.8$ in HgTe), 
reflecting the significantly larger effective mass of the heavy hole compared to the light hole.

This unusual feature---where the two branches of surface states are neither fully connected nor fully disconnected---can be understood as follows. 
We incorporate the nodal line as a non-abelian element (of dimension two) within the geometric connection scheme linking the two branches. 
To achieve this, we introduce a connection matrix (comparator)
$U_{\text{nl}}(2,\bar{2})$ 
that connects the branches across the nodal line, defined by the relations
$U_{\text{nl}}(2,\bar{2})\left|\chi_{2}\right\rangle=\left|\chi_{\bar{2}}\right\rangle$ and $U_{\text{nl}}(\bar{2},2)\left|\chi_{\bar{2}}\right\rangle=\left|\chi_{2}\right\rangle$.
These equations are complemented by the following unitarity condition: $U_{\text{nl}}(2,\bar{2})U_{\text{nl}}(\bar{2},2)=1$.
In the basis $\left\{\chi_{2},\chi_{\bar{2}}\right\}$, we find
\begin{equation}
U_{\text{nl}}(2,\bar{2})=\left(\begin{array}{cc}
\frac{\gamma_1}{2\gamma_2}&-\eta_+\eta_-\\
\eta_+\eta_- & \frac{\gamma_1}{2\gamma_2}
\end{array}\right)\equiv \left[U_{\text{nl}}(\bar{2},2)\right]^\dagger.
\label{eqU22b}
\end{equation}
Analogous to Eq.~(\ref{eqdisconmeasure}), the matrix in  Eq.~(\ref{eqU22b}) also shows that the two branches are fully disconnected when $\gamma_1=0$ 
and perfectly connected when $\left|\gamma_1\right|= 2\gamma_2$.
Note that $U_{\text{nl}}(2,\bar{2})$ has no overlap with the space of bulk states other than those which form the nodal line.
We therefore conclude that, in general, the two surface state branches are intermediately connected via the nodal line, 
which acts as a degenerate subspace enabling a continuous though non-abelian connection to either branch.

In other words, the two surface-state branches are connected in a non-abelian way through the nodal line, which acts as a two-dimensional degenerate subspace.

\subsection{Van Hove singularities and projected density of states}
\label{secPDOS}

The bulk spectrum can also be obtained analytically, in the presence of spin-orbital terms, for the special directions of $\bm{k}=(\bm{q},k_z)$ 
lying in the planes $(1\bar{1}0)$ and $(110)$.
The spin-orbit interaction lifts the two-fold degeneracy, resulting in a projected density of states shown in Fig.~\ref{fig:ProjDOSSemiMetal}(b).
For the two bands which repel away from the nodal line [gray tones for positive $q$ in Fig.~\ref{fig:ProjDOSSemiMetal}(b)],  
we find
\begin{equation}
E_{1,\pm}^{\text{bulk}}=\gamma_1\left(k_z^2+q^2\right)-\frac{3}{2}\alpha q
\pm\sqrt{3\left(k_z^2+\frac{q^2}{4}\right)(\alpha+2\gamma_3 q)^2+4\gamma_2^2(k_z^2-k_{\Delta}^2)^2},
\label{eqE1bulkpm}
\end{equation}
where $k_{\Delta}$ is defined in Eq.~(\ref{eqkDeltaAppx}).
The van Hove singularities obtained from this spectrum are similar to those 
in Eqs.~(\ref{eqEvH0}) and~(\ref{eqEvH1}), 
except that they are generalized here to include the spin-orbital terms containing $\alpha$,
\begin{equation}
E_{1,\pm}^{\text{vH0}}=\gamma_1q^2-\frac{3}{2}\alpha q
\pm\sqrt{\frac{3}{4}(\alpha+2\gamma_3 q)^2q^2+4\gamma_2^2k_{\Delta}^4},
\label{eqE1vH0}
\end{equation}
\begin{equation}
\begin{split}
E_{1,\pm}^{\text{vH1}} &=-\frac{\gamma_1\Delta}{4\gamma_2}-\frac{3}{2}\alpha q
+\frac{3}{2}\gamma_1\left[q^2-\frac{\left(\alpha+2\gamma_3 q\right)^2}{4\gamma_2^2}\right] \\
&\pm\frac{\sqrt{3}}{2}\eta_+\eta_-\left(\alpha+2\gamma_3q\right) 
\sqrt{q^2-\frac{3\left(\alpha+2\gamma_3q\right)^2}{4\gamma_2^2}+4k_\Delta^2}.
\end{split}
\label{eqE1vH1}
\end{equation}
The existence condition for the van Hove singularity vH1 reads
\begin{equation}
\gamma_1\left(E_{1,\pm}^{\text{vH1}} +\frac{3\alpha q}{2}\right)<-\gamma_2\Delta+\left(\gamma_1^2+2\gamma_2^2\right)q^2
-\frac{3}{2}\left(\alpha+2\gamma_3q\right)^2.
\label{eqE1vH1exist}
\end{equation}
The band edge of the projected density of states is given by vH1, provided vH1 exists; otherwise it is given by vH0.

The remaining two bands that form the nodal line are denoted by $E_{2,\pm}^{\text{bulk}}(q)$.
An expression for $E_{2,\pm}^{\text{bulk}}$ is obtained from the right-hand side of Eq.~(\ref{eqE1bulkpm})
via the substitution  $q\to -q$.
Their van Hove singularities within the inner region (inside the nodal line) are also obtained by this same substitution applied to 
Eqs.~(\ref{eqE1vH0})-(\ref{eqE1vH1exist}).
However, for the outer region, we must additionally apply the replacement
$\eta_-\to -\eta_-$ in Eq.~(\ref{eqE1vH1}), 
following the same rule used for the surface states.

\begin{figure*}[!ht]
\includegraphics[width=0.48\textwidth]{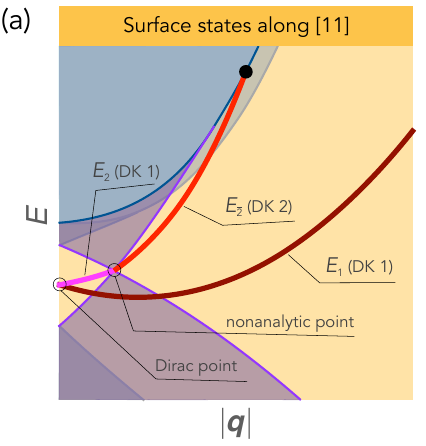}
\hfill
\includegraphics[width=0.48\textwidth]{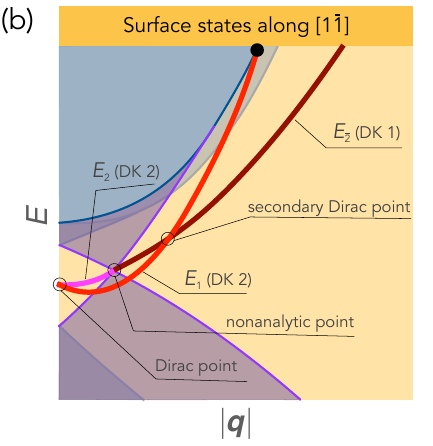}
\caption{\label{SurfaceStatesExact0}
Surface states along two high-symmetry directions of the wave vector $\bm{q}$.
(a) Case of $\bm{q}\parallel \left[11\right]$: 
The lower surface-state branch, $E_1(q)$, 
is identified as the DK 1 branch in Fig.~\ref{fig:ProjDOSSemiMetal}(a).
It is shifted to the right and deformed by the spin-orbit interaction.
The upper branch, $E_2(q)$, begins as a DK 1 branch that is shifted to the left,
forming a 2D Dirac point at $q=0$.
At the nodal line, it changes non-analytically into the branch
$E_{\bar{2}}(q)$, which is identified as a DK 2 branch.
(b) Case of $\bm{q}\parallel \left[1\bar{1}\right]$: 
Here, the lower branch
$E_1(q)$ corresponds to the DK 2 branch in  Fig.~\ref{fig:ProjDOSSemiMetal}(a)
and is similarly shifted to the right and deformed.
The upper branch $E_2(q)$ initially appears as a DK 2
but undergoes a transition to a DK 1 branch at the nodal line.
Outside of the nodal-line, $E_{\bar{2}}(q)$ intersects $E_1(q)$, 
forming a secondary Dirac point at finite $q$ in addition to the one at $q=0$.
}
\end{figure*}

In Fig.~\ref{SurfaceStatesExact0}, we illustrate the complex structure of the PDOS and the surface states arising from the spin-orbit interaction.
It is important to realize that, because of the spin splitting, surface states are allowed to cross regions of PDOS without coupling to the bulk states.
This is possible because of the orthogonality due to ``spin'' preserved by the time-reversal symmetry as a distinct quantum number.
If spin scattering is absent, the discrete surface state branches can be resolved on the background of the PDOS without broadening.

In Fig.~\ref{SurfaceStatesExact0}, we illustrate the complex structure of the PDOS 
and the associated surface states that arise from the spin-orbit interaction. 
A crucial point is that the spin splitting allows these surface states to traverse regions of non-zero PDOS without coupling to bulk states. 
This decoupling is possible due to the orthogonality of the distinct pseudospin channels---a quantum number preserved by time-reversal symmetry in this system. 
In the absence of spin-scattering processes, the discrete surface-state branches can therefore be resolved clearly against the PDOS background without broadening.

For the high-symmetry directions, we can identify each surface state branch with one of the two DK branches 
shown in Fig.~\ref{fig:ProjDOSSemiMetal}(a). 
Specifically, for $s=1$ [see Fig.~\ref{SurfaceStatesExact0}(a)],
the lower branch $E_1(q)$ is a DK 1 branch,
which can be verified by following this state continuously to the limit $\alpha\to 0$.
The upper branch transitions from a DK 1 to a DK 2 branch at the nodal line. These surface states form a two-dimensional Dirac cone at
 $q=0$, consistent with the results of \cite{Kibis}.

For $s=-1$ [see Fig.~\ref{SurfaceStatesExact0}(b)], 
the lower branch $E_1(q)$ is a DK 2 branch.
The upper branch changes from a DK 2 to a DK1 branch at the nodal line.
In addition to the primary Dirac point at
$q=0$, a secondary Dirac point appears along this direction in the region outside the nodal line.

Overall, the spin-orbit interaction does not radically alter the surface states. 
As expected for a linear-in-momentum term, its primary effects are shifts and deformations of the energy branches.
The most significant consequences, however, are the non-analytic behavior of the surface states at the nodal line and the emergence of two secondary Dirac points along the diagonal
$k_x+(\alpha/\left|\alpha\right|)k_y=0$ outside the nodal loop.

\section{Conclusions}
We have developed a comprehensive theory of surface states across the strain-induced topological phase diagram of inverted-band semiconductors like HgTe. 
Using a minimal Luttinger model extended by strain and spin-orbit terms, we obtained exact analytical solutions that reveal several key phenomena. 
The most striking is a topologically enforced non-analyticity in the surface-state dispersion at the projection of the bulk nodal line, accompanied by an abrupt spinor reorientation. 
This behavior demonstrates that the surface states are connected in a singular, non-abelian fashion through the bulk degeneracy.

Our analysis unifies the description of surface states in these systems, showing how Dyakonov-Khaetskii states evolve continuously 
under strain and are further deformed by spin-orbit coupling. 
We predict the emergence of unique features such as secondary Dirac points and a momentum-space patch structure for the surface states. 
Furthermore, we established a clear hierarchy of energy scales governing the transitions between Dirac, nodal-line, and Weyl semimetal phases, 
which delineates the experimental resolution needed to probe each regime.

These findings advance the fundamental understanding of how bulk topology manifests in the surface electronic structure of gapless topological materials. 
The identified signatures provide specific targets for experimental detection in strained heterostructures, 
paving the way for the exploration of novel topological transport phenomena and responses in this versatile class of tunable semiconductors.


\section*{Acknowledgements}
A.K.\ gratefully acknowledges the hospitality of the Donostia International Physics Center (DIPC) and the Centro de F\'{i}sica de Materiales (CFM-MPC), Centro Mixto CSIC-UPV/EHU, during his scientific visit.

\paragraph{Author contributions}
V.N.G. and A.K. contributed equally to this work.

\paragraph{Funding information}
V.N.G. acknowledges funding from the Spanish MCIN/AEI/10.13039/501100011033 
through grant PID2024-160189NA-I00, 
co-funded by the European Regional Development Fund (FEDER, EU), 
as well as from the Basque Government through grants PIBA\_2025\_1\_0018, 
IT-1591-22, and IKUR/RESONANT.

\bibliography{references_SurfaceStates}

\end{document}